\documentclass[icol, referee, pdflatex, sn-aps]{sn-jnl}

\usepackage{graphicx}%
\usepackage{multirow}%
\usepackage{amsmath,amssymb,amsfonts}%
\usepackage{amsthm}%
\usepackage{mathrsfs}%
\usepackage[title]{appendix}%
\usepackage{xcolor}%
\usepackage{textcomp}%
\usepackage{manyfoot}%
\usepackage{booktabs}%
\usepackage{algorithm}%
\usepackage{algorithmicx}%
\usepackage{algpseudocode}%
\usepackage{listings}%

\theoremstyle{thmstyleone}%
\theoremstyle{thmstyletwo}%

\theoremstyle{thmstylethree}%

\begin{document}

\title[Article Title]{Synergistic effects of rare-earth doping on the magnetic properties of orthochromates: A machine learning approach}




\author[1]{\fnm{} \sur{Guanping Xu}}
\equalcont{These authors contributed equally to this work.}
\author[1]{\fnm{} \sur{Zirui Zhao}}
\equalcont{These authors contributed equally to this work.}
\author[1]{\fnm{} \sur{Muqing Su}}
\author*[1]{\sur{Hai-Feng Li}}\email{haifengli@um.edu.mo}
\affil[1]{\orgdiv{Institute of Applied Physics and Materials Engineering}, \orgname{University of Macau}, \orgaddress{\street{Avenida da Universidade}, \city{Taipa}, \postcode{999078}, \state{Macao SAR}, \country{China}}}


\abstract{Multiferroic materials, particularly rare-earth orthochromates (RECrO$_3$), have garnered significant interest due to their unique magnetic and electric-polar properties, making them promising candidates for multifunctional devices. Although extensive research has been conducted on their antiferromagnetic (AFM) transition temperature (N$\acute{\textrm{e}}$el temperature, $T_\textrm{N}$), ferroelectricity, and piezoelectricity, the effects of doping and substitution of rare-earth (RE) elements on these properties remain insufficiently explored. In this study, convolutional neural networks (CNNs) were employed to predict and analyze the physical properties of RECrO$_3$ compounds under various doping scenarios. Experimental and literature data were integrated to train machine learning models, enabling accurate predictions of $T_\textrm{N}$, besides remanent polarization ($P_\textrm{r}$) and piezoelectric coefficients ($d_{33}$). The results indicate that doping with specific RE elements significantly impacts $T_\textrm{N}$, with optimal doping levels identified for enhanced performance. Furthermore, high-entropy RECrO$_3$ compounds were systematically analyzed, demonstrating how the inclusion of multiple RE elements influences magnetic properties. This work establishes a robust framework for predicting and optimizing the properties of RECrO$_3$ materials, offering valuable insights into their potential applications in energy storage and sensor technologies.}




\maketitle

\section{Introduction}

Multiferroic materials have garnered significant research interest due to their simultaneous magnetic and electric-polar properties \cite{1,2,3,4,5,6}. Specifically, rare-earth (RE) orthochromate compounds, RECrO$_3$ (RE = rare-earth elements), a class of perovskite-structured oxides, exhibit multiferroic behavior \cite{7,8,9,10,11,12,13}, including piezoelectric, magnetic, and optoelectronic properties \cite{14,15,16,17,18,19,20}. These diverse properties can be tailored by varying the type and proportion of RE elements. Such tunability has made these materials highly attractive for applications in storage-type and self-powered multifunctional devices, thermistors, and magnetic cooling devices \cite{21,22,23,24,25}.

The interaction between RE and Cr sites is hypothesized to induce spin-phonon coupling, magnetostriction effects, and negative magnetization behavior. These phenomena provide valuable insights into the mechanisms underlying ferroelectricity and other intriguing properties of RECrO$_3$ compounds \cite{26,27,28}. The antiferromagnetic (AFM) transition temperature, or N$\acute{\textrm{e}}$el temperature ($T_\textrm{N}$), is strongly connected to the physical and chemical properties of multiferroic materials and is significantly influenced by RE ions and specific site doping. Despite significant progress in exploring the AFM transition temperature of RE orthochromates and their ferroelectric and piezoelectric properties, the origin of ferroelectricity in orthochromates remains elusive.

Many studies have examined how different RE$^{3+}$ ions influence the properties of RECrO$_3$ \cite{29,30,31,32}, focusing on changes in the superexchange interaction within the RE$^{3+}$ sublattice and the overlap of $t_{\textrm{2g}}$-$e_\textrm{g}$ orbitals, which affect the electric polarization and piezoelectricity of these compounds \cite{33,34,35,36,37}. However, in-depth analyses of the effects of RE ions, as well as doping or substituting other elements in RE orthochromates, are still lacking. Limited information exists on the effects of doping or substituting elements such as Pr$^{3+}$ or Ce$^{3+}$ (which replace part of the RE$^{3+}$ ions) on the properties of these materials. For instance, compounds such as (La$_{1-x}$Ce$_x$)CrO$_3$ or (La$_{1-x}$Pr$_x$)CrO$_3$ could exhibit significant changes in $T_\textrm{N}$, ferroelectricity, and magnetic properties. Similarly, doping at the Cr site would also affect $T_\textrm{N}$, as well as the ferroelectric and magnetic properties of RECrO$_3$ compounds \cite{38,39,40,41,42,43,44,45}.

Machine learning has emerged as a powerful tool to enhance research efficiency, particularly in the fields of nanomaterials, piezoelectrics, and power batteries. It is especially valuable for material screening and performance optimization. Convolutional neural networks (CNNs), a typical machine learning method, have proven instrumental in processing and analyzing vast datasets, enabling accurate predictions in material screening and providing guidance for the synthesis of new materials. In recent years, machine learning has been extensively applied in physics and chemistry \cite{46,47,48,49,50,51,52,53}. However, no reports currently exist on its application to RE orthochromates (RECrO$_3$), nor on how doping at RE sites influences the N$\acute{\textrm{e}}$el temperature $T_\textrm{N}$, piezoelectric coefficient $d_{33}$, and remanent polarization $P_\textrm{r}$. This gap highlights the need for research on how these factors alter the magnetic and ferroelectric properties of these materials \cite{54,55,56}.

To address this gap, this study investigates the magnetic properties of RE orthochromates (RECrO$_3$), focusing on the effects of substituting and doping different RE elements at specific sites. We systematically examine variations in the types, amounts, and proportions of RE elements to align with practical applications. Additionally, we analyze the N$\acute{\textrm{e}}$el temperature ($T_\textrm{N}$) in doped RE orthochromate compounds, with a particular focus on the relationship between magnetism and RE-doping elements. The goal is to predict and screen compounds with optimal performance and perform comparative verification. Furthermore, this study explores the potential applications of RE orthochromates, particularly in energy storage and sensor design, emphasizing their versatility in advanced technologies.

To achieve this, we propose a method to predict and screen RE orthochromate materials with suitable N$\acute{\textrm{e}}$el temperatures using CNNs \cite{57,58,59}. Our approach begins by screening various RE elements (e.g., La, Ce, Pr, Yb) and collecting experimental and simulation data related to doping elements, doping sites, and other relevant parameters. The acquired dataset is then preprocessed to extract key features that affect the AFM transition temperature ($T_\textrm{N}$). Compared to traditional methods such as molecular dynamics and density functional theory (DFT), which demand significant computational resources and rely heavily on potential energy functions, machine learning algorithms offer notable advantages. They can utilize large datasets to achieve high accuracy and efficiency in predicting how doping behavior affects macroscopic performance, without requiring complex recalculations \cite{60,61,62}.

We subsequently build a CNN model for predicting $T_\textrm{N}$, which visualizes the screening process of RECrO$_3$ doped with various elements. The model architecture consists of multiple graph convolutional layers for feature aggregation and updating, along with optional graph attention layers to capture complex relationships between doping elements and sites \cite{63}. Using the preprocessed data, we train and validate the CNN model to accurately predict the factors influencing the N$\acute{\textrm{e}}$el temperature, piezoelectric coefficient, and multiferroic performance. These include doping elements, doping sites, and other potential methodologies. This approach not only provides insights into the mechanisms affecting the N$\acute{\textrm{e}}$el temperature of RE orthochromate materials but also offers a practical and efficient tool for screening and synthesizing RECrO$_3$ materials with enhanced performance.

\section{Computational methods and models}

CNNs are a type of deep learning model extensively utilized in the fields of image processing and feature extraction \cite{64}. These networks are adept at capturing local features in data through their convolution operations, demonstrating superior performance when dealing with high-dimensional data \cite{65,66}. In our previous work, we have successfully developed CNN models to: (i) Optimize doping strategies for enhancing the electrochemical performance and structural integrity of ternary nickel–cobalt–manganese cathode materials in lithium-ion batteries \cite{ZHAO2024112982}. (ii) Efficiently evaluate the ionic conductivity and electrochemical properties of ion-doped NASICON materials \cite{Zhao2024-1}. (iii) Understand and predict interface diffusion phenomena in applied materials \cite{63}. This section will elucidate the fundamental principles of CNNs and provide their mathematical expressions.

\subsection{Fundamental principles} 

In understanding the fundamental principles of CNNs, it is crucial to examine the core components of the network. The convolutional layer is the primary feature extractor, utilizing convolution operations to capture local features. The convolution operation can be formally expressed as:
\begin{eqnarray}
(f * g)(t) = \int_{-\infty}^{\infty} f(\tau)g(t - \tau) d\tau.
\label{1}
\end{eqnarray}
In the discrete case, the convolution operation simplifies to:
\begin{eqnarray}
(f * g)[n] = \sum_{m=-\infty}^{\infty} f[m]g[n - m],
\label{2}
\end{eqnarray}
where \( f \) is the input signal, \( g \) is the convolution kernel (or filter), and \( n \) is the index of the output signal. In CNNs, the convolution kernel is typically a small matrix that extracts local features by performing dot product operations with local regions of the input data. Following the convolutional layer is typically a pooling layer, whose purpose is to reduce dimensionality and computational load while retaining important features. Common pooling operations include max pooling and average pooling. The mathematical expression for max pooling is:
\begin{eqnarray}
P_{max}(x) = \max_{i \in R} (x_i),
\label{3}
\end{eqnarray}
and for average pooling:
\begin{eqnarray}
P_{avg}(x) = \frac{1}{|R|} \sum_{i \in R} x_i,
\label{4}
\end{eqnarray}
where \( R \) is the size of the pooling window, and \( x_i \) are the elements within the window.

The final layers of a CNN are usually fully connected layers, similar to traditional multilayer perceptrons. These layers map inputs to outputs through linear transformations and activation functions. This process can be mathematically expressed as:
\begin{eqnarray}
y = \sigma(Wx + b),
\label{5}
\end{eqnarray}
where \( W \) is the weight matrix, \( x \) is the input vector, \( b \) is the bias vector, and \( \sigma \) is the activation function, such as rectified linear unit (ReLU):
\begin{eqnarray}
\sigma(x) = \max(0, x).
\label{6}
\end{eqnarray}

By combining these layers, CNNs can effectively extract and learn hierarchical features from data. The training process of CNNs involves forward propagation, which calculates the network output, and back propagation, which updates network parameters using the gradient descent method to minimize the loss function. Common forms of the loss function \( L \) include mean squared error (MSE) and cross-entropy (CE):
\begin{eqnarray}
L_{MSE} = \frac{1}{N} \sum_{i=1}^{N} (y_i - \hat{y_i})^2, \textrm{and}
\label{7}
\end{eqnarray}
\begin{eqnarray}
L_{CE} = -\sum_{i=1}^{N} y_i \log(\hat{y_i}),
\label{8}
\end{eqnarray}
where \( y_i \) is the true value, \( \hat{y_i} \) is the predicted value, and \( N \) is the number of samples.

CNNs are highly effective in processing and predicting complex data, such as the magnetic properties of RECrO$_3$ compounds \cite{67, Qahtan_2024}. This efficiency stems from the combined operation of convolutional, pooling, and fully connected layers. A key strength of CNNs lies in their ability to automatically extract features from data, making them particularly promising for applications in materials science. However, to maximize the potential of CNNs, preparation of a high-quality training dataset is essential.

\subsection{Dataset preparation and data preprocessing} 

High-quality datasets are essential for applying CNNs to predict the magnetic and piezoelectric properties of RECrO$_3$ materials \cite{67, Qahtan_2024}. A carefully curated dataset can significantly improve both the predictive accuracy and the generalization capability of the model. The following sections detail the data sources, preprocessing steps, and feature selection process essential for effective CNN training.

\begin{figure} [!t]
\centering
\includegraphics[width=0.68\textwidth]{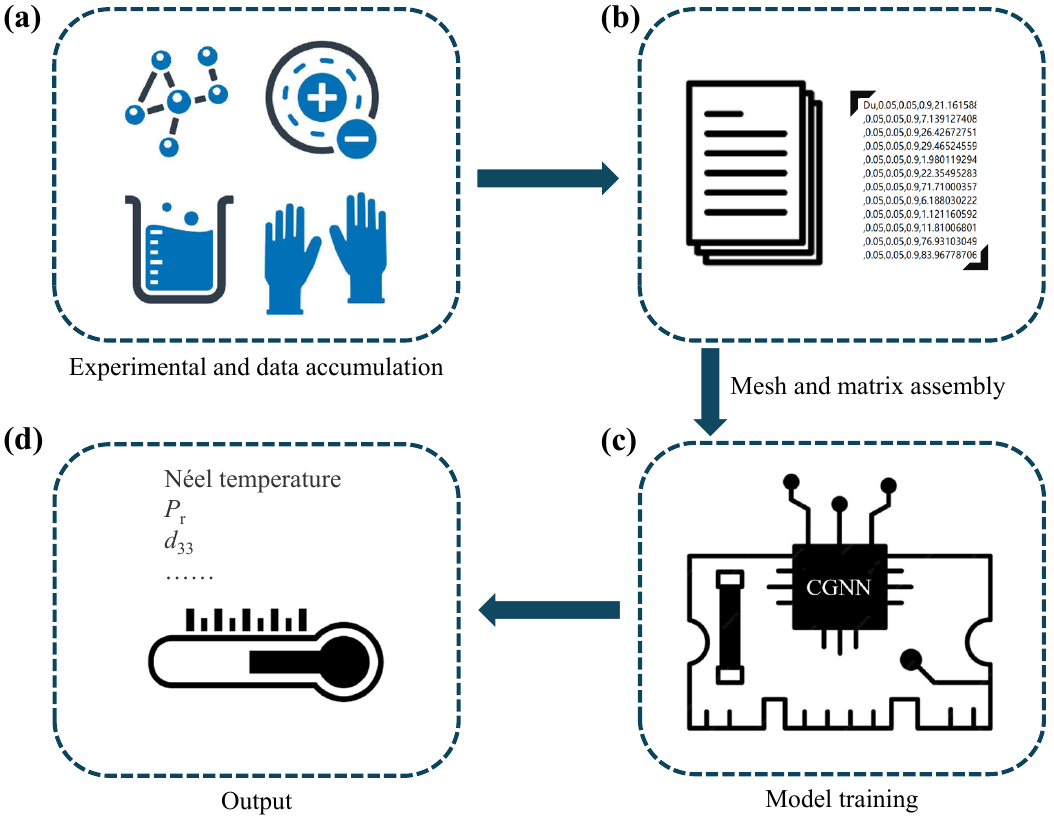}
\caption{Workflow for predicting the physical properties of the RECrO$_3$ compound family. \textbf{a} We compiled data from prior studies in the literature, \textbf{b} which was then integrated into a convolutional neural network model, \textbf{c} subsequently trained, and the model's performance evaluated for the stability and accuracy of the predicted results. \textbf{d} These include parameters such as the N$\acute{\textrm{e}}$el temperature ($T_\textrm{N}$), remanent electric polarization ($P_\textrm{r}$), and piezoelectric coefficient ($d_{33}$).}
\label{Fig1}
\end{figure}

As shown in Fig.~\ref{Fig1}, the workflow for predicting the physically relevant properties of the RECrO$_3$ family of compounds \cite{67, Qahtan_2024} is systematically implemented. The primary data sources include experimental data and literature data. Experimental data is obtained through controlled experiments measuring the magnetic, piezoelectric, and ferroelectric properties of RECrO$_3$ materials \cite{67, Qahtan_2024} under various conditions. Although highly reliable, this data is costly and time-intensive to acquire. In contrast, literature data is extracted from published research papers and databases. This data is more extensive and cost-effective but may suffer from issues related to quality and consistency.

Experimental data typically includes measurements such as the N$\acute{\textrm{e}}$el temperature, remanent polarization, piezoelectric coefficients, and lattice parameters under varying environmental conditions (e.g., temperature, pressure) \cite{40, 67}. Literature data is gathered through a comprehensive review of scientific publications and data mining from relevant databases.

Raw data often requires preprocessing to make it suitable for training CNNs. Key preprocessing steps include data cleaning, normalization, and augmentation to ensure data quality and consistency. Data cleaning involves removing noise and outliers from the dataset to improve accuracy and reliability. Common methods include deleting missing values or using interpolation techniques to fill gaps. Normalization scales the data to a specific range (e.g., 0 to 1) to eliminate differences in the dimensions of various features. Standard normalization techniques, such as min-max normalization and \emph{Z}-score standardization, are commonly applied.

\begin{eqnarray}
x' = \frac{x - \min(x)}{\max(x) - \min(x)}, \textrm{and}
\label{9}
\end{eqnarray}
\begin{eqnarray}
z = \frac{x - \mu}{\sigma},
\label{10}
\end{eqnarray}
where \( \min(x) \) and \( \max(x) \) are the minimum and maximum values of the feature, respectively, \( \mu \) is the mean, and \( \sigma \) is the standard deviation. Data augmentation generates additional training samples by transforming the existing data (e.g., rotation, scaling, translation) to enhance the model's generalization ability.

After dataset preparation, selecting the most influential features is crucial for the model's predictive performance.

\subsection{Feature selection} 

Feature selection is crucial in predicting the magnetic, piezoelectric and ferroelectric properties of RECrO$_3$ materials \cite{67, Qahtan_2024}, aiming to identify the most influential variables for the model input. Key factors affecting these properties include chemical composition, crystal structure, lattice constants, density, and electronic structure, which can be obtained through experimental measurements or computational simulations.

Feature engineering involves transforming and combining raw features to generate new, more representative features. For instance, principal component analysis can be used to reduce the dimensionality of high-dimensional features:
\begin{eqnarray}
Z = XW,
\label{11}
\end{eqnarray}
where \( X \) is the original feature matrix, \( W \) is the matrix of eigenvectors, and \( Z \) is the reduced feature matrix. Common feature selection methods include correlation analysis, recursive feature elimination, and LASSO regression, which effectively identify features that significantly impact the model's predictive performance.

The construction of a high-quality training dataset requires meticulous preparation, including data sourcing, preprocessing, and feature selection. This robust foundation is essential for effectively training and optimizing CNNs, which, in turn, enhances the accuracy of predicting the properties of RECrO$_3$ materials \cite{67, Qahtan_2024}.

\subsection{Model construction and training} 

Constructing an effective CNN model and training it are essential for predicting the properties of RECrO$_3$ materials. The CNN model designed in this study consists of six convolutional layers, six pooling layers, and twelve fully connected layers. The following sections describe the architecture and functions of each component, as well as the training process and related challenges.

The designed CNN model includes six convolutional layers that extract local features from the input data. Each convolutional layer's output is processed non-linearly using the ReLU activation function. The convolution operation is described by Equ.~\ref{2}. Following each convolutional layer, six pooling layers are employed to reduce data dimensionality and computational load. Max pooling operations are performed as described in Equ.~\ref{3}. The model also incorporates twelve fully connected layers to map high-dimensional features to the output space. Each fully connected layer performs linear transformations combined with the ReLU activation function (Equ.~\ref{5}). Finally, the output layer generates the prediction results using appropriate activation functions, such as softmax or linear activation.

After defining the model architecture, the next step is to train the model effectively to achieve accurate predictions. The training process involves dataset division, hyperparameter tuning, and optimization methods. The dataset is split into training, validation, and test sets: the training set trains the model, the validation set tunes hyperparameters and selects the model, and the test set evaluates its performance. Key hyperparameters, such as learning rate, batch size, and number of iterations, are carefully selected. Common optimization algorithms include stochastic gradient descent and its variants (e.g., Adam optimizer). The loss function quantifies the error between predicted and true values, with common examples being MSE (Equ.~\ref{7}) and cross-entropy (Equ.~\ref{8}).

The training process further includes forward and back propagations. Forward propagation calculates the network's output, while back propagation updates the network parameters using gradient descent to minimize the loss function. The mathematical formulations for forward and back propagations are provided in Equs.~\ref{9} and \ref{10}, respectively.
\begin{eqnarray}
\text{Forward: } a^{(l)} = \sigma(W^{(l)} a^{(l-1)} + b^{(l)}), \textrm{and}
\label{17}
\end{eqnarray}
\begin{eqnarray}
\text{Backward: } \frac{\partial L}{\partial W^{(l)}} = \frac{\partial L}{\partial a^{(l)}} \cdot \frac{\partial a^{(l)}}{\partial W^{(l)}},
\label{18}
\end{eqnarray}
where \( l \) denotes the \( l \)-th layer, \( a \) is the activation value, \( W \) and \( b \) are the weights and biases, respectively, and \( L \) is the loss function.

During training, issues such as overfitting, vanishing gradients, or exploding gradients may arise. Common solutions to mitigate overfitting include adding regularization terms (e.g., \emph{L}2 regularization), using dropout layers, and employing data augmentation techniques. To address vanishing or exploding gradients, methods such as appropriate initialization techniques (e.g., He initialization or Xavier initialization) and gradient clipping are applied. These techniques ensure the stability of the training process, laying the foundation for reliable model performance.

Building upon these techniques, a meticulously designed and trained CNN model facilitates the efficient prediction of the properties of RECrO$_3$ materials \cite{67, Qahtan_2024}, providing robust support for advancements in materials science. Once the model training is complete, its performance must undergo rigorous evaluation to ensure reliability and accuracy. Key evaluation metrics, such as accuracy and recall, are employed to validate the model's effectiveness.

\subsection{Model evaluation and results analysis} 

Model evaluation is a critical step in assessing the performance of a CNN model for predicting the properties of RECrO$_3$ materials \cite{67, Qahtan_2024}. In this section, we will discuss the evaluation metrics, present the results, and analyze the findings.

To evaluate the performance of the CNN, we employed several standard metrics including accuracy, precision, recall, and \emph{F}1-score. Accuracy is a key metric that indicates the proportion of correct predictions made by the model. It is calculated as follows:
\begin{eqnarray}
\text{Accuracy} = \frac{\text{Number of Correct Predictions}}{\text{Total Number of Predictions}}.
\label{19}
\end{eqnarray}

Precision, recall, and \emph{F}1-score are particularly useful for assessing the model's performance on imbalanced datasets. Precision measures the proportion of true positive predictions among all positive predictions, while recall measures the proportion of true positive predictions among all actual positives. The \emph{F}1-score is the harmonic mean of precision and recall, providing a balanced measure of the two:
\begin{eqnarray}
\text{Precision} = \frac{\text{True Positives}}{(\text{True Positives}) + (\text{False Positives})},
\label{20}
\end{eqnarray}
\begin{eqnarray}
\text{Recall} = \frac{\text{True Positives}}{(\text{True Positives}) + (\text{False Negatives})}, \textrm{and}
\label{21}
\end{eqnarray}
\begin{eqnarray}
F\text{1-score} = 2 \cdot \frac{\text{Precision} \cdot \text{Recall}}{\text{Precision} + \text{Recall}}.
\label{22}
\end{eqnarray}

\section{Results and discussion}

\subsection{Prediction and comparison of magnetic properties of RECrO$_3$}

Utilizing a type-boosting classifier, we developed a systematic machine learning approach to predict the N$\acute{\textrm{e}}$el temperature of materials with rare-earth chromate perovskite structures \cite{68,69,70,71,72,73}. We analyzed data from thousands of RECrO$_3$ compounds, considering the types of doped elements, doping sites, and other relevant features \cite{74,75,76,77,78}. This analysis enabled us to establish a predictive model that closely aligns with the $T_\textrm{N}$ trends reported in the literature. The following sections discuss the alignment of our predictions with experimental data and highlight areas for further refinement.

\begin{figure} [!t]
\centering
\includegraphics[width=0.68\textwidth]{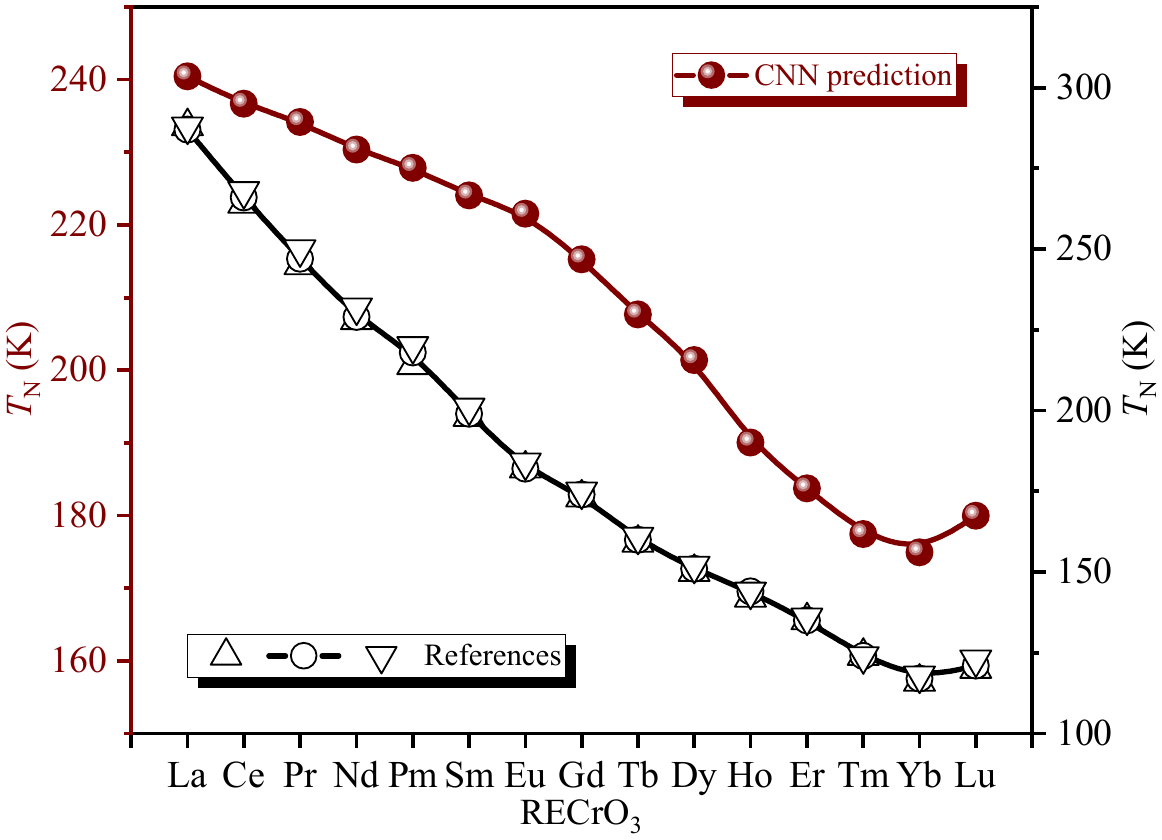}
\caption{(Left) Predicted AFM transition temperature ($T_\textrm{N}$) as a function of RE elements for RECrO$3$ orthochromates, as obtained from a convolutional neural network model. (Right) Summary of $T_\textrm{N}$ values for RECrO$_3$ compounds as reported in the literature.}
\label{Fig2}
\end{figure}

Our study on the AFM transition temperature ($T_\textrm{N}$) trends in rare-earth chromates (RECrO$_3$) demonstrated close alignment between our model's predictions and the empirical $T_\textrm{N}$ trends, as shown in Fig.~\ref{Fig2}. DFT calculations further support these trends \cite{79, 80}, yielding predicted $T_\textrm{N}$ values that approximate empirical data. However, both the CNN model and DFT calculations consistently deviate from the actual $T_\textrm{N}$ values by approximately 2–60 K. For instance: (i) The experimental N$\acute{\textrm{e}}$el temperature ($T^{\textrm{exp}}_\textrm{N}$) of LaCrO$_3$ is 288 K \cite{PhysRevLett.106.057201}, which is 47.56 K higher than the CNN-predicted value ($T^{\textrm{CNN}}_\textrm{N}$ = 240.44 K). (ii) YbCrO$_3$ has an experimental $T^{\textrm{exp}}_\textrm{N}$ of 117 K, lower than the predicted $T^{\textrm{CNN}}_\textrm{N}$ of 174.92 K.
(iii) In contrast, NdCrO$_3$ has an experimental N$\acute{\textrm{e}}$el temperature of 228 K, which closely matches the predicted $T^{\textrm{CNN}}_\textrm{N}$ of 230.36 K. These discrepancies suggest that the current model requires refinement, particularly in accounting for the effects of RE elements with 4\emph{f} electrons. Nevertheless, our qualitative CNN model successfully predicts the $T_\textrm{N}$ trend of RECrO$_3$ compounds.

Previously, we reported that as the ionic radius of lanthanide rare-earth elements decreases from europium (Eu$^{3+}$) to lutetium (Lu$^{3+}$), the lattice parameters (\emph{a}, \emph{b}, and \emph{c}) and unit-cell volume (\emph{V}) show an almost linear relationship with the ionic radius, with Lu$^{3+}$ being an exception \cite{40, 80}. Analysis of the downward trend in the RECrO$_3$ series illustrated in Fig.~\ref{Fig2} reveals that a similar linear relationship holds for the AFM transition temperature $T_\textrm{N}$ versus ionic radius for RE ions ranging from La$^{3+}$ to Yb$^{3+}$. However, for the Lu$^{3+}$ ion, the $T_\textrm{N}$ of LuCrO$_3$ does not follow this linear relationship. This deviation can be attributed to the non-magnetic nature of Lu$^{3+}$ due to its fully filled 4$f^{14}$ orbitals. Consequently, this also leads to the absence of potential Cr$^{3+}$ (3\emph{d})-Lu$^{3+}$ (4\emph{f}) coupling in the LuCrO$_3$ compound. Interestingly, our predictions suggest that the $T_\textrm{N}$ (179.96 K) of LuCrO$_3$ is slightly higher than that (174.92 K) of YbCrO$_3$, despite the Yb$^{3+}$ ion having a marginally larger ionic radius and being magnetic (4$f^{13}$).

\begin{figure} [!t]
\centering
\includegraphics[width=0.68\textwidth]{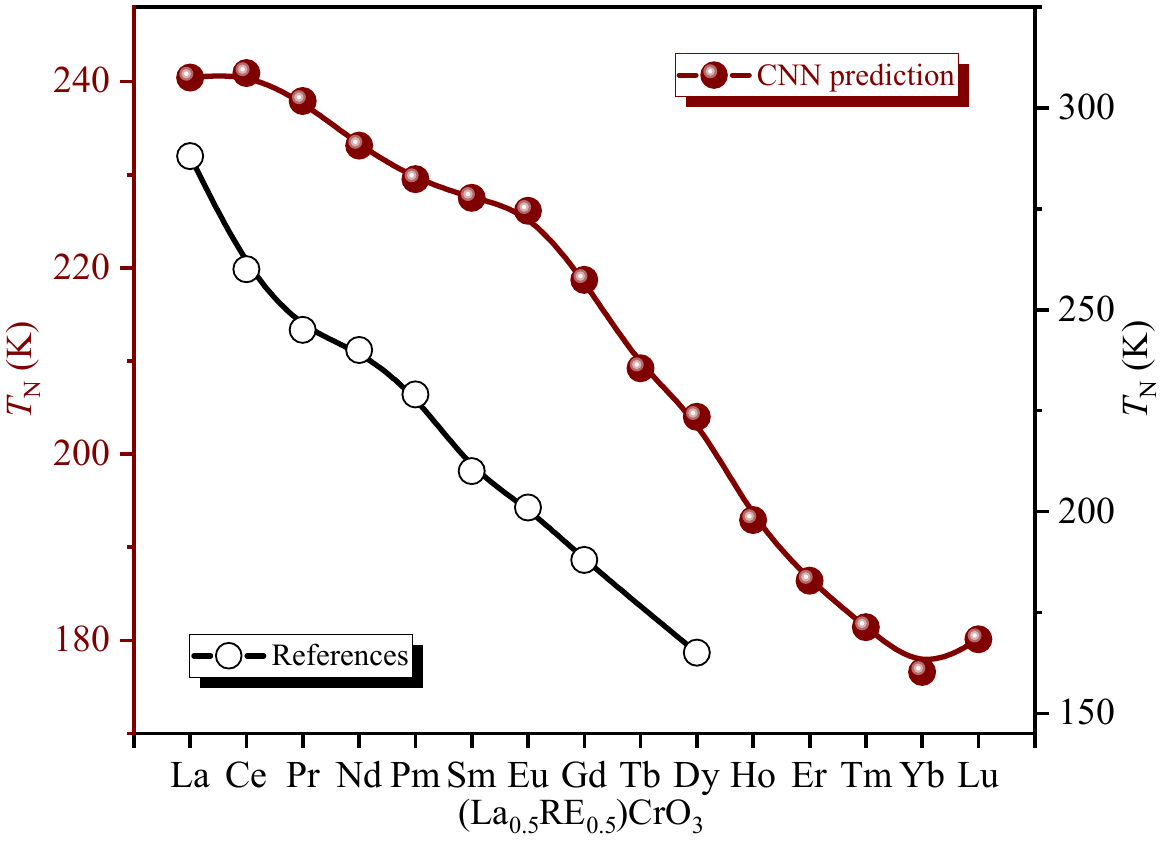}
\caption{(Left) Predicted AFM transition temperature ($T_\textrm{N}$) as a function of RE elements for (La$_{0.5}$RE$_{0.5}$)CrO$_3$ orthochromates, as obtained from a convolutional neural network model. (Right) Comparison of $T_\textrm{N}$ values for (La$_{0.5}$RE$_{0.5}$)CrO$_3$ compounds summarized from the literature.}
\label{Fig3}
\end{figure}

In alignment with our model design and predictive framework, we have categorized the related compounds into three distinct types based on their doping sites and levels, as well as the number of doping elements for high-entropy RECrO$_3$-based materials:

(i) (La$_{1-\textrm{x}}$RE$\textrm{x}$)CrO$_3$. In this type of compound, we systematically analyze the AFM properties by adjusting the doping of RE elements during the machining learning process. Using our trained and validated CNN model, we predict the $T_\textrm{N}$ trend for the (La$_{0.5}$RE$_{0.5}$)CrO$_3$ compound, as illustrated in Fig.~\ref{Fig3}, and determine the optimal doping level \emph{x} (Fig.~\ref{Fig4} and Table~\ref{Table 1}).

(ii) (La$_{1-x-y}$RE1$_x$RE2$_y$)CrO$_3$, where RE1 and RE2 represent the doping RE elements. By optimizing the doping of RE1 and RE2 elements at the La position, we schematically generate the corresponding orthochromate compounds with a chemical formula (La$_{0.5}$RE1$_{0.25}$RE2$_{0.25}$)CrO$_3$ (Fig.~\ref{Fig5}) and evaluate the $T_\textrm{N}$ trend of orthochromates with double-element doping (Table~\ref{Table 2}).

(iii) Our investigation into high entropy compounds involved:
\begin{itemize}
\item La$_{1/2}$RE0$_0$($\sum_{i=1}^{14}$RE\textit{i}$_{1/28}$)CrO$_3$, where RE0$_0$ indicates zero doping for the RE0 element.
\item La$_{1/2}$($\sum_{i=1}^{15}$RE\textit{i}$_{1/30}$)CrO$_3$, where RE$\textit{i}$ = Sc, Lu, Ce, Pr, Nd, Pm, Sm, Eu, Gd, Tb, Dy, Ho, Er, Tm, and Yb.
\end{itemize}

We selected La$_{1/2}$($\sum$RE\textit{i})$_{1/2}$CrO$_3$ as the parent compound and doped it with 14 or 15 RE\textit{i} elements to create high-entropy compounds with multiple elements. The results related to these compounds were presented in Fig.~\ref{Fig6} and Table~\ref{Table 3}.

\subsection{Single rare-earth element doping in LaCrO$_3$} 

Following the successful prediction of the AFM transition temperature ($T_\textrm{N}$) trends for RE orthochromates RECrO$_3$ using our established model, we extended our approach to the RE-doped orthochromate system. We selected LaCrO$_3$ as the parent compound for doping due to its highest N$\acute{\textrm{e}}$el temperature among all RE orthochromates \cite{PhysRevLett.106.057201} and introduced various RE elements at different doping ratios to form La$_{1-x}$RE$_x$CrO$_3$ compounds. To facilitate comparison, we focused on La$_{0.5}$RE$_{0.5}$CrO$_3$ compounds for model predictions, as illustrated in Fig.~\ref{Fig3}. Notably, many single-element-doped LaCrO$_3$ compounds remain under-explored, leading to incomplete data in the literature regarding all possible doping scenarios. Nevertheless, the existing $T\textrm{N}$ data generally align with our predictions.

\begin{figure} [!t]
\centering
\includegraphics[width=0.68\textwidth]{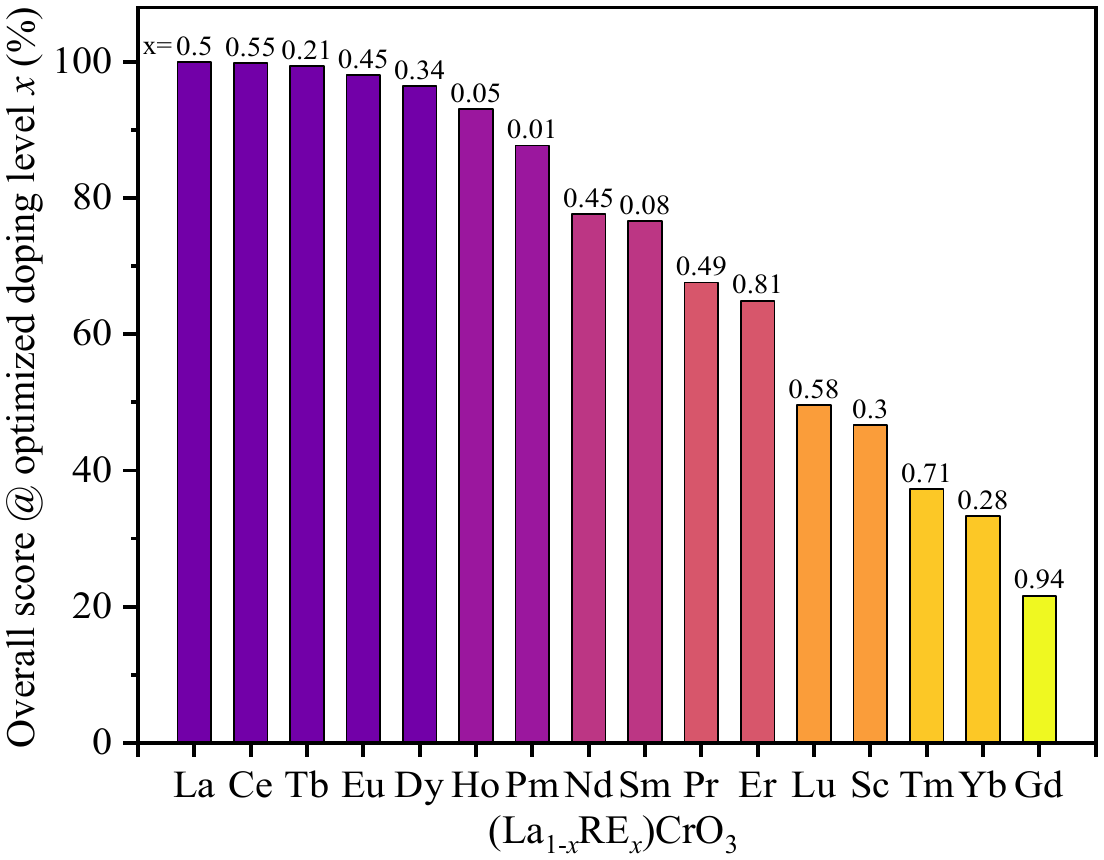}
\caption{Predicted performance metrics of the AFM transition temperature ($T_\textrm{N}$) as a function of doping with various RE elements in (La$_{1-x}$RE$_x$)CrO$_3$ orthochromates at the optimal doping concentration (\emph{x}), as calculated using a convolutional neural network model.}
\label{Fig4}
\end{figure}

\begin{figure} [!t]
\centering
\includegraphics[width=0.68\textwidth]{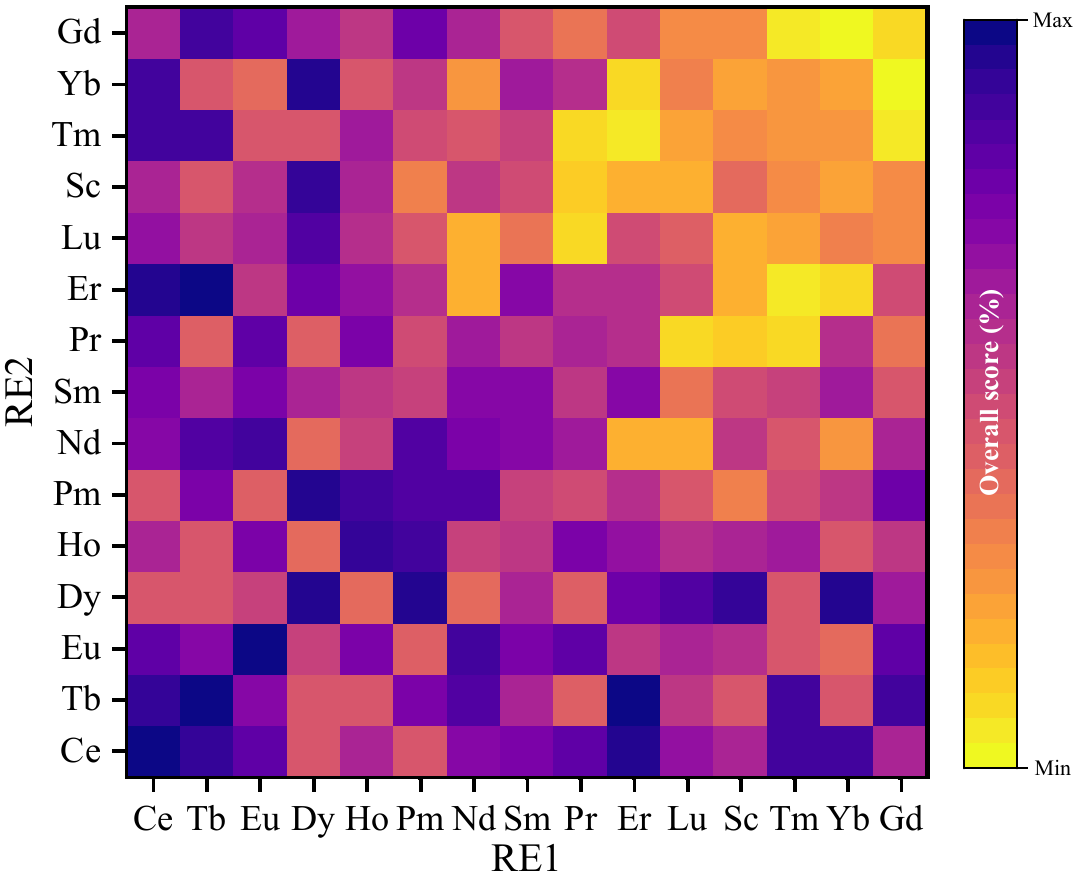}
\caption{Predicted overall performance metrics of the AFM transition temperature ($T_\textrm{N}$) as a function of co-doping with two RE elements (RE1 and RE2) for the (La$_{0.5}$RE1$_{0.25}$RE2$_{0.25}$)CrO$_3$ orthochromate, as calculated using a convolutional neural network model.}
\label{Fig5}
\end{figure}

Based on the predictions from our model, we further explored the doping effect of different RE elements on LaCrO$_3$. For example, the N$\acute{\textrm{e}}$el temperature of pure LaCrO$_3$ (denoted as La$_{0.5}$La$_{0.5}$CrO$_3$) is 288 K \cite{PhysRevLett.106.057201}. However, doping with 0.45 atm\% Nd reduces the $T_\textrm{N}$ of (La$_{0.55}$Nd$_{0.45}$)CrO$_3$ to 218.69 K. This reduction is primarily attributed to lattice shrinkage and the resultant alterations in spin interactions caused by the incorporation of Ce$^{3+}$ ions. Similar trends are observed with other dopants, where lattice deformation contributes to a decrease in the $T_\textrm{N}$ value. In addition to the 0.5 atm\% doping ratio, our established model can predict various doping ratios for La$_{1-x}$RE$_x$CrO$_3$ compounds. By comprehensively analyzing the RE doping effect, we can determine the $T_\textrm{N}$ values for these compounds and identify the optimal doping ratio corresponding to the highest $T_\textrm{N}$ (Fig.~\ref{Fig4}).

Through the analysis of different doping ratios, Fig.~\ref{Fig4} and Table~\ref{Table 1} present the predicted maximum $T_\textrm{N}$ values and the corresponding optimal doping ratios. The results show that the $T_\textrm{N}$ of undoped LaCrO$_3$ remains the highest, supporting the physical principle that doping induces lattice distortion, which reduces $T_\textrm{N}$ \cite{80}. Furthermore, the impact on $T_\textrm{N}$ is notably dependent on the nature of the doping elements. For instance, doping with 0.01 atm\% Pm results in a higher $T_\textrm{N}$ (224.17 K) compared to doping with 0.45 atm\% Nd ($T_\textrm{N}$ = 218.69 K), while doping with Gd at 0.94 atm\% produces the lowest $T_\textrm{N}$ (176.91 K). Specifically, the $T_\textrm{N}$ value of La$_{0.04}$Gd$_{0.94}$CrO$_3$ is minimized at this high optimal doping ratio. Our model predictions indicate that binary chromate compounds with tailored $T_\textrm{N}$ values can be synthesized by adjusting the doping ratios to meet the requirements of specific experimental and application contexts.

\subsection{Two rare-earth elements doping in LaCrO$_3$}

After our model successfully predicted the $T_\textrm{N}$ trends and specific values for single-element RE orthochromates RECrO$_3$ and single RE-doped (La$_{1-x}$RE$_x$)CrO$_3$ orthochromates, we extended our analysis to double RE doping using LaCrO$_3$ as the parent material. We introduced two RE elements, RE1 and RE2 (which may be identical or different), to create doubly doped (La$_{0.5}$RE1$_{0.25}$RE2$_{0.25}$)CrO$_3$ compounds. Predictions were made based on the structural features and performance indicators of the selected elements. Fig.~\ref{Fig5} illustrates a confusion matrix that visualizes the best predictions across all experimental data. The confusion matrix, a standard tool for evaluating binary classification performance, consists of fifteen rows and fifteen columns, where the rows represent one doping element (RE1) and the columns represent another doping element (RE2). Correct predictions are located on the diagonal of the matrix, which is divided into four quadrants.

Based on the predictions from our model, we further explored the doping effects of different RE elements on LaCrO$_3$. The overall performance scores, exceeding 80\%, along with the calculated AFM transition temperatures ($T_\textrm{N}$) for the co-doping model with RE1 and RE2 elements, are summarized in Table~\ref{Table 2}. For example, the (La$_{0.5}$Ce$_{0.25}$Ce$_{0.25}$)CrO$_3$ compound achieves the highest overall performance score of 99.88\% with a maximum $T_\textrm{N}$ of 244.63 K. For comparison, the performance score is 89.32\% for (La$_{0.5}$Eu$_{0.25}$\\Nd$_{0.25}$)CrO$_3$ with $T_\textrm{N}$ = 215.02 K, and 84.01\% for (La$_{0.5}$Ce$_{0.25}$Pr$_{0.25}$)CrO$_3$ with $T_\textrm{N}$ = 198.89 K. Our model achieves a prediction accuracy of 92\%, demonstrating robust performance in predicting $T_\textrm{N}$ variations in RE orthochromates.

\subsection{Prediction of variations in the $T_\textrm{N}$ for high entropy orthochromates}

\begin{figure*} [!t]
\centering
\includegraphics[width=0.82\textwidth]{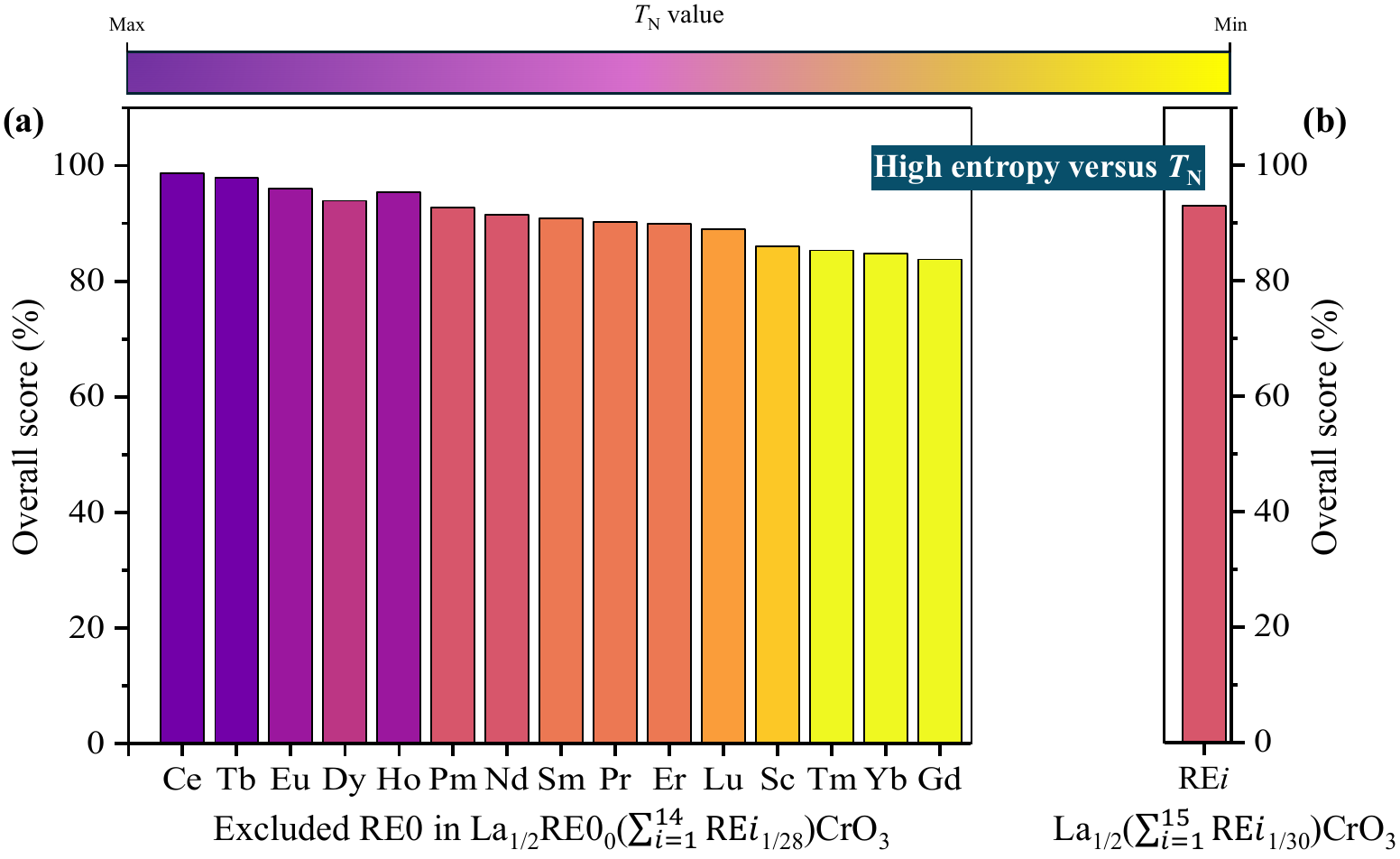}
\caption{\textbf{a} Predicted overall performance metrics of the AFM transition temperature ($T_\textrm{N}$) as a function of excluding an RE0 element for La$_{1/2}$RE0$_0$($\sum_{i=1}^{14}$RE\textit{i}$_{1/28}$)CrO$_3$ (RE = rare-earth elements) in high-entropy rare-earth chromate compounds co-doped with RE elements. \textbf{b} Predicted overall performance metric of the AFM transition temperature ($T_\textrm{N}$) as a function of all RE\emph{i} (\emph{i} = 1--15) elements for La$_{1/2}$($\sum_{i=1}^{15}$RE\textit{i}$_{1/30}$)CrO$_3$ in the high-entropy RE chromate compound. Both sets of results were calculated using a convolutional neural network model.}
\label{Fig6}
\end{figure*}

Regarding the change in the AFM transition temperature ($T_\textrm{N}$) of high-entropy RE orthochromates, we selected La$_{1/2}$($\sum$RE\textit{i})$_{1/2}$CrO$_3$ as the parent compound and 15 different RE$\textit{i}$ ions (Sc, Lu, Ce, Pr, Nd, Pm, Sm, Eu, Gd, Tb, Dy, Ho, Er, Tm, and Yb) as doping elements. The results obtained from our CNNs model are presented in Fig.~\ref{Fig6} and summarized in Table~\ref{Table 3}. The high-entropy orthochromate doped with all 15 RE$\textit{i}$ ions (Fig.~\ref{Fig6}b), i.e., La$_{1/2}$(Sc$_{1/30}$Lu$_{1/30}$Ce$_{1/30}$Pr$_{1/30}$Nd$_{1/30}$Pm$_{1/30}$Sm$_{1/30}$Eu$_{1/30}$Gd$_{1/30}$\\Tb$_{1/30}$Dy$_{1/30}$Ho$_{1/30}$Er$_{1/30}$Tm$_{1/30}$Yb$_{1/30}$)\\CrO$_3$, achieves a $T_\textrm{N}$ value of 240.41 K with an overall performance score of 93.05\%. Interestingly, when Ce$^{3+}$ ions are excluded (i.e., doping with 14 RE$\textit{i}$ ions), the resulting high-entropy orthochromate \\La$_{1/2}$(Sc$_{1/28}$Lu$_{1/28}$Pr$_{1/28}$Nd$_{1/28}$Pm$_{1/28}$Sm$_{1/28}$Eu$_{1/28}$Gd$_{1/28}$Tb$_{1/28}$Dy$_{1/28}$Ho$_{1/28}$Er$_{1/28}$Tm$_{1/28}$Yb$_{1/28}$)CrO$_3$ exhibits the highest $T_\textrm{N}$ value of 242.15 K. In contrast, excluding Yb$^{3+}$ reduces the $T_\textrm{N}$ value to 180.71 K, as shown in Fig.~\ref{Fig6}a and Table~\ref{Table 3}.

Based on these results, we further examined the impact of RE elements on the magnetic properties of high-entropy orthochromates. For simplicity, the doping ratio of each RE$\textit{i}$ ion was fixed in this study. However, there may exist an optimal doping level for individual elements that could achieve even higher $T_\textrm{N}$ values. Our findings also indicate that increasing the number of doping elements does not necessarily improve the magnetic properties of orthochromates. The selection of specific doping elements and determining their optimal doping levels are critical factors in advancing high-entropy materials research.

\subsection{Prediction and comparison of piezoelectric and ferroelectric properties of RECrO$_3$} 

In addition to forecasting the $T_\textrm{N}$ trends of RECrO$_3$ orthochromates and related compounds, we also predicted the piezoelectric and ferroelectric performance parameters of RE orthochromates, including the remnant polarization ($P\textrm{r}$) and piezoelectric coefficient ($d_{33}$). These parameters are critical for evaluating the practical applicability of materials, particularly in electronic devices and sensors.

\begin{figure} [!t]
\centering
\includegraphics[width=0.68\textwidth]{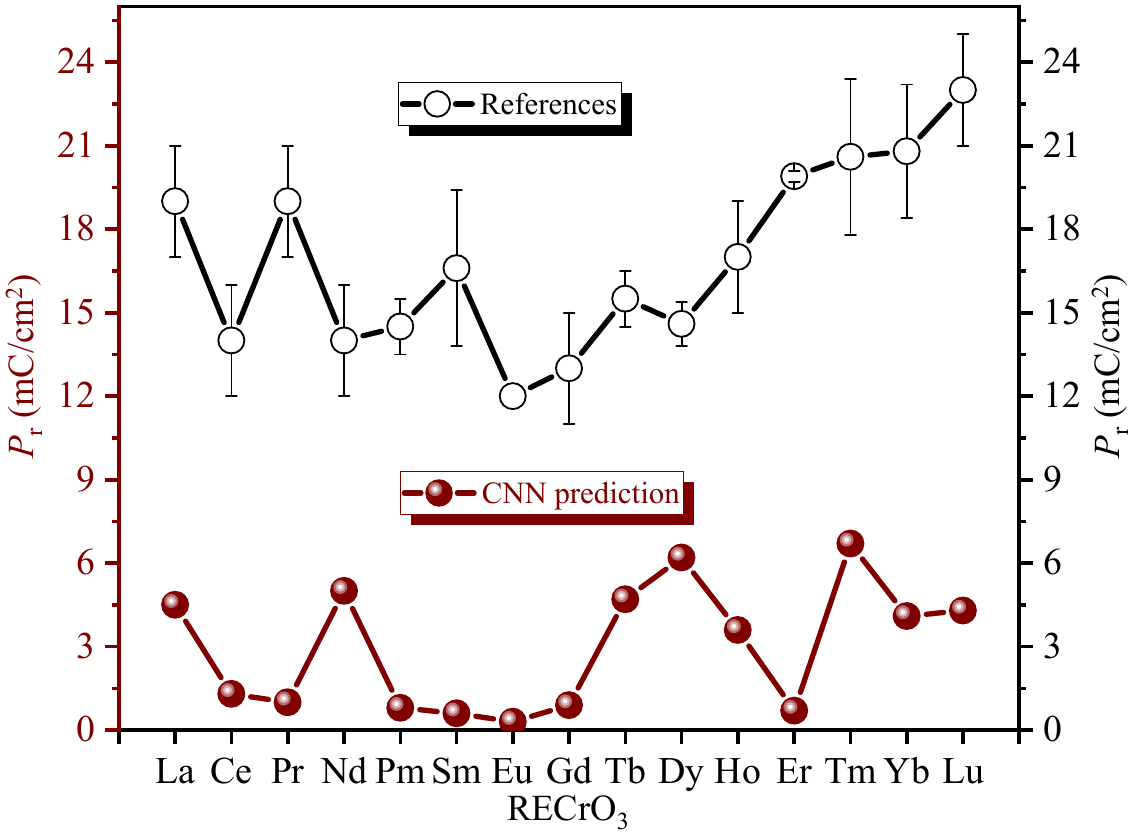}
\caption{(Left) Predicted values of remanent electric polarization ($P_\textrm{r}$) as a function of doping with RE elements for RECrO$_3$ orthochromates, as calculated using a convolutional neural network model. (Right) Comparison of $P_\textrm{r}$ values for RECrO$_3$ compounds compiled from the literature.}
\label{Fig7}
\end{figure}

\begin{figure} [!t]
\centering
\includegraphics[width=0.68\textwidth]{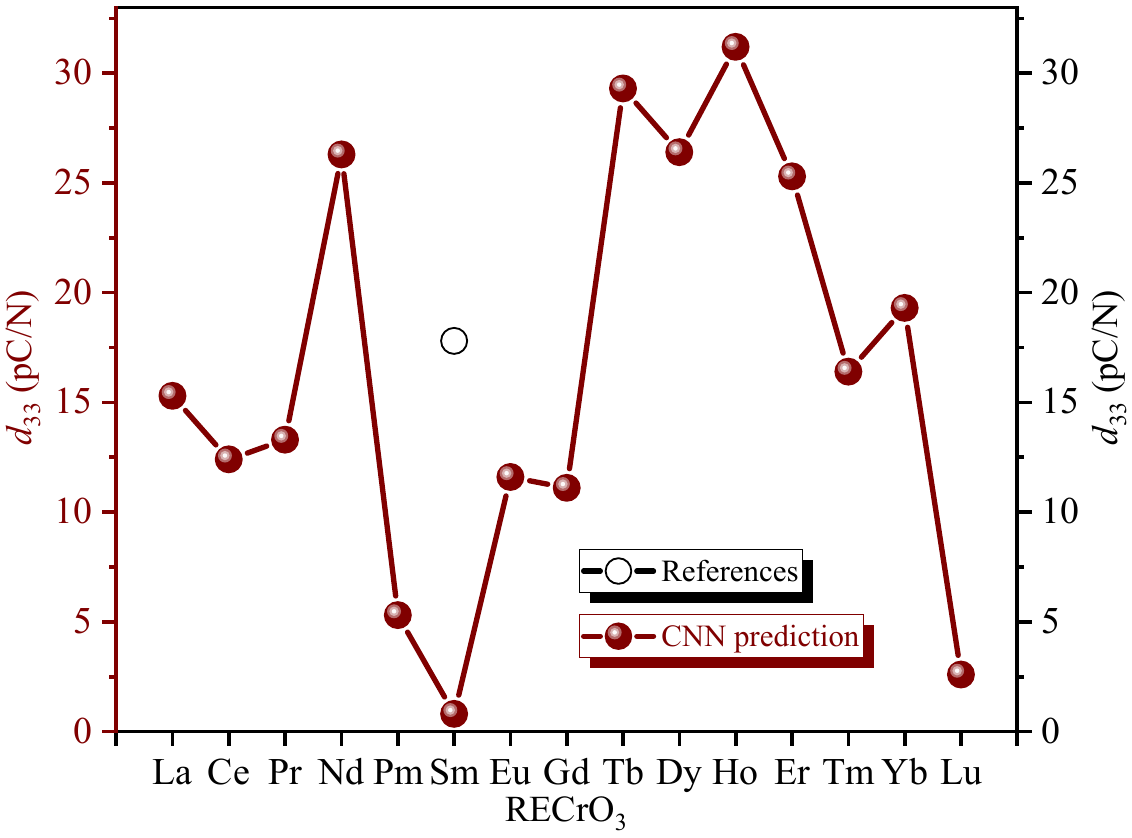}
\caption{(Left) Predicted values of piezoelectric coefficient ($d_{33}$) as a function of doping with RE elements for RECrO$_3$ orthochromates, calculated using a convolutional neural network model. (Right) Summary of $d_{33}$ values for RECrO$_3$ compounds reported in the literature.}
\label{Fig8}
\end{figure}

To validate these predictions, we have plotted Figs.~\ref{Fig7} and \ref{Fig8}, which illustrate the trends of $P_\textrm{r}$ (Fig.~\ref{Fig7}) and $d_{33}$ (Fig.~\ref{Fig8}) in RE orthochromates. Fig.~\ref{Fig7} shows the remnant polarization ($P_\textrm{r}$), a commonly used ferroelectric performance parameter. While the trend predicted by our model aligns roughly with experimental data reported in the literature, we observed that these materials exhibit low $P_\textrm{r}$ values. Additionally, we analyzed the piezoelectric coefficient ($d_{33}$), another critical performance parameter. The $d_{33}$ values predicted by our model, as presented in Fig.~\ref{Fig8}, reveal that the piezoelectric coefficients of RECrO$_3$ materials are relatively small and significantly lower than those of most polymer materials \cite{qian2023fluoropolymer}.

Based on these results, we further evaluated the practical applications of RE orthochromates. Our analysis indicates that there are few related studies on their applications, suggesting limited potential and scope for these materials in practical use.

\subsection{Verification section}

To validate the accuracy and efficiency of the CNN models we developed, we selected experimental data from various orthochromate systems that were not included in our collected database. This approach ensures the broad applicability and robustness of the models.
\begin{itemize}
\item Using our models, we first predicted the properties of three undoped RECrO$_3$ materials, such as the N$\acute{\textrm{e}}$el temperature ($T_\textrm{N}$), achieving an accuracy of up to 92\%.
\item Next, we randomly selected three single-element-doped (La$_{1-x}$RE$_x$)CrO$_3$ materials to verify both the trends and numerical values of $T_\textrm{N}$, attaining similarly high accuracy.
\item We then tested three types of dual-element-doped (La$_{1-x-y}$RE1$_x$RE2$_y$)CrO$_3$ materials, ensuring efficient validation of the model's predictions.
\item Finally, we validated the model using high-entropy materials incorporating multiple RE elements. These tests revealed variations in $T_\textrm{N}$ and provided valuable guidance for synthesizing high-entropy compounds in multi-element systems, thereby enhancing our understanding of the fundamental theory of high-entropy materials.
\end{itemize}

\section{Conclusions}

This study presents a systematic investigation into the physical properties of RE orthochromates (RECrO$_3$), focusing on the influence of RE doping and substitution on their AFM transition temperature (N$\acute{\textrm{e}}$el temperature, $T_\textrm{N}$), as well as their ferroelectric and piezoelectric properties. By leveraging CNN models, we integrate experimental data with literature-reported results to predict and analyze the behavior of RECrO$_3$ compounds under various doping scenarios. The computational approach enables accurate predictions of $T_\textrm{N}$, remanent polarization ($P_\textrm{r}$), and piezoelectric coefficients ($d_{33}$), providing a comprehensive framework for understanding the effects of RE doping on these materials.

Key findings from the study reveal that specific RE doping levels significantly influence $T_\textrm{N}$, with optimal doping concentrations identified for enhanced magnetic properties. Single-element doping demonstrated that the ionic radius of RE ions and the lattice shrinkage of the unit cell play a critical role in modulating magnetic interactions. In contrast, co-doping with multiple RE elements highlighted synergistic effects that can further optimize $T_\textrm{N}$. The analysis of high-entropy RECrO$_3$ compounds showed that while multiple RE elements introduce complex interactions, not all combinations improve material performance, emphasizing the need for targeted doping strategies.

Additionally, the study confirms that the piezoelectric and ferroelectric properties of RECrO$_3$ remain relatively modest, with limited applicability in these domains compared to other piezoelectric materials. This work establishes a robust machine learning-based methodology for predicting and optimizing the physical properties of RECrO$_3$ materials. The findings bridge existing gaps in understanding the interplay between RE doping and material properties, offering valuable insights for the design of advanced multifunctional materials. Furthermore, the study provides a foundation for future experimental and computational exploration of RE orthochromates, advancing their potential applications.

\clearpage

\backmatter

\section*{Acknowledgments}

This work was supported by the Science and Technology Development Fund, Macao SAR (File Nos. 0104/2024/AFJ, 0002/2024/TFP, and 0115/2024/RIB2), University of Macau (MYRG-GRG2024-00158-IAPME and MYRG-GRG2025-00251-IAPME), and the Guangdong-Hong Kong-Macao Joint Laboratory for Neutron Scattering Science and Technology (Grant No. 2019B121205003). 

\section*{Authors’ contributions}

Guanping Xu and Zirui Zhao contributed equally. 
Guanping Xu: Conceptualization, data curation, formal analysis, investigation, visualization, writing-original draft. 
Zirui Zhao: Conceptualization, data curation, formal analysis, investigation, visualization, writing-original draft. 
Muqing Su: Formal analysis, investigation, methodology, visualization. 
Hai-Feng Li: Conceptualization, funding acquisition, methodology, project administration, supervision, visualization, writing-review \& editing.

\section*{Availability of data and materials}

All data generated or analyzed during this study are included in this published article.

\section*{Declarations}

\textbf{Ethics approval and consent to participate}

The authors declare they have upheld the integrity of the scientific record.

\textbf{Consent for publication}

The authors give their consent for publication of this article.

\textbf{Competing interests}

The authors declare that they have no competing interests.

\textbf{Non-financial interests}

Hai-Feng Li serves as an editor for the AAPPS Bulletin

\textbf{Financial interests}

The authors declare they have no financial interests.

\clearpage

\bibliographystyle{sn-aps}
\bibliography{Xu-2.bib}

\clearpage

\clearpage

\clearpage

\clearpage

\clearpage

\clearpage

\clearpage

\clearpage

\clearpage

\begin{table}[!t]
\centering
\renewcommand*{\thetable}{\Roman{table}}
\caption{Predicted values of AFM transition temperature ($T_\textrm{N}$) as a function of doping with RE elements for (La$_{1-x}$RE$_x$)CrO$_3$ orthochromates at the optimal doping concentration (\emph{x}), as calculated using a convolutional neural network model. The data presented in this table correspond to Fig.~\ref{Fig4}.}
\label{Table 1}
\setlength{\tabcolsep}{3.8mm}{}
\renewcommand{\arraystretch}{1.1}
\begin{tabular}{ccc}
\hline
Doped elements          & Optimized doping level           & $T_\textrm{N}$               \\
                        & (x)                              & (K)                          \\
\hline
La                      & 0.50                             & 244.63                       \\
Ce                      & 0.55                             & 240.36                       \\
Tb                      & 0.21                             & 235.11                       \\
Eu                      & 0.45                             & 232.61                       \\
Dy                      & 0.34                             & 229.67                       \\
Ho                      & 0.05                             & 226.46                       \\
Pm                      & 0.01                             & 224.17                       \\
Nd                      & 0.45                             & 218.69                       \\
Sm                      & 0.08                             & 208.34                       \\
Pr                      & 0.49                             & 202.16                       \\
Er                      & 0.81                             & 192.38                       \\
Lu                      & 0.58                             & 189.26                       \\
Sc                      & 0.30                             & 185.53                       \\
Tm                      & 0.71                             & 181.36                       \\
Yb                      & 0.28                             & 180.39                       \\
Gd                      & 0.94                             & 176.91                       \\
\hline
\end{tabular}
\end{table}

\clearpage

\begin{table}[!h]
\centering
\renewcommand*{\thetable}{\Roman{table}}
\caption{Predicted values of AFM transition temperature ($T_\textrm{N}$) as a function of co-doping with two RE elements (RE1 and RE2) for the (La$_{0.5}$RE1$_{0.25}$RE2$_{0.25}$)CrO$_3$ orthochromate, as calculated using a convolutional neural network model with an overall performance score greater than 80\%. The data presented in this table correspond to Fig.~\ref{Fig5}.}
\label{Table 2}
\setlength{\tabcolsep}{5.8mm}{}
\renewcommand{\arraystretch}{1.1}
\begin{tabular}{cccc}
\hline
RE1        & RE2                      & Scores                    & $T_\textrm{N}$           \\
           &                          & (\%)                      & (K)                      \\
\hline
Ce         & Ce                       & 99.88                     & 244.63                   \\
Dy         & Pm                       & 95.09                     & 233.86                   \\
Tb         & Er                       & 98.31                     & 233.37                   \\
Pm         & Pm                       & 87.71                     & 229.67                   \\
Eu         & Eu                       & 98.07                     & 224.17                   \\
Dy         & Sc                       & 91.52                     & 223.34                   \\
Ce         & Er                       & 96.58                     & 223.10                   \\
Tb         & Gd                       & 91.40                     & 219.48                   \\
Dy         & Yb                       & 95.11                     & 218.49                   \\
Ce         & Tm                       & 89.69                     & 218.45                   \\
Tb         & Tm                       & 91.49                     & 215.76                   \\
Eu         & Nd                       & 89.32                     & 215.02                   \\
Ce         & Tb                       & 94.04                     & 213.56                   \\
Dy         & Lu                       & 86.98                     & 213.56                   \\
Ce         & Yb                       & 89.42                     & 213.36                   \\
Dy         & Er                       & 80.93                     & 209.48                   \\
Eu         & Gd                       & 83.37                     & 208.76                   \\
Pm         & Nd                       & 86.30                     & 208.69                   \\
Tb         & Nd                       & 86.43                     & 208.42                   \\
Tb         & Tb                       & 99.36                     & 208.34                   \\
Ho         & Pm                       & 88.92                     & 205.73                   \\
Ce         & Eu                       & 83.96                     & 203.53                   \\
Dy         & Dy                       & 96.47                     & 202.16                   \\
Eu         & Pr                       & 83.58                     & 202.06                   \\
Pm         & Gd                       & 80.79                     & 201.33                   \\
Ce         & Pr                       & 84.01                     & 198.89                   \\
Ho         & Ho                       & 93.03                     & 192.38                   \\
\hline
\end{tabular}
\end{table}

\clearpage

\begin{table}[!t]
\centering
\renewcommand*{\thetable}{\Roman{table}}
\caption{Predicted values of AFM transition temperature ($T_\textrm{N}$) as a function of excluding an RE0 element for La$_{1/2}$RE0$_0$($\sum_{i=1}^{14}$RE\textit{i}$_{1/28}$)CrO$_3$ (RE = rare-earth elements) in high-entropy rare-earth chromate compounds co-doped with RE\emph{i} (\emph{i} = 1--14) elements. For comparison, we also show the $T_\textrm{N}$ value of the high-entropy La$_{1/2}$($\sum_{i=1}^{15}$RE\textit{i}$_{1/30}$)CrO$_3$ chromate compound doped with all 15 RE\emph{i} elements. Both sets of results were calculated using a convolutional neural network model. The data presented in this table correspond to Fig.~\ref{Fig6}.}
\label{Table 3}
\setlength{\tabcolsep}{3.8mm}{}
\renewcommand{\arraystretch}{1.1}
\begin{tabular}{lll}
\hline
                                       & Scores                  & $T_\textrm{N}$            \\
                                       & (\%)                    & (K)                       \\
\hline
Doping 15 RE\emph{i}                   & 93.05                   & 240.41                    \\
\hline
Doping 14 RE\emph{i}, excluding RE0    &                         &                           \\
Ce (i.e., RE0)                         & 98.69                   & 242.15                    \\
Tb                                     & 97.82                   & 240.21                    \\
Eu                                     & 95.99                   & 235.29                    \\
Ho                                     & 95.42                   & 233.69                    \\
Dy                                     & 93.91                   & 233.22                    \\
Pm                                     & 92.80                   & 229.94                    \\
Nd                                     & 91.55                   & 225.95                    \\
Sm                                     & 90.81                   & 219.61                    \\
Pr                                     & 90.19                   & 213.68                    \\
Er                                     & 89.93                   & 206.55                    \\
Lu                                     & 88.97                   & 194.13                    \\
Sc                                     & 85.97                   & 189.44                    \\
Tm                                     & 85.32                   & 185.56                    \\
Gd                                     & 83.75                   & 184.72                    \\
Yb                                     & 84.77                   & 180.71                    \\
\hline
\end{tabular}
\end{table}

\clearpage

\end{document}